\documentclass[twocolumn,preprintnumbers,amsmath,amssymb,prl,superscriptaddress,longbibliography]{revtex4-1}

\usepackage{subfigure}
\usepackage{graphicx}
\usepackage{siunitx}
\usepackage{hyperref}
\usepackage{soul}
\usepackage[normalem]{ulem}
\hypersetup{colorlinks=true, citecolor=blue, urlcolor=blue, linkcolor=blue}
\allowdisplaybreaks

\begin{document}

\title{Tunable surface plasmons in Weyl semimetals TaAs and NbAs}

\author{Gennaro Chiarello}
\email{gennaro.chiarello@fis.unical.it} \affiliation{Department of
Physics, University of Calabria, via ponte Bucci, 31/C, 87036
Rende (CS), Italy}

\author{Johannes Hofmann}
\email{jbh38@cam.ac.uk}
\affiliation{Department of Applied Mathematics and Theoretical Physics, University of Cambridge, Centre for Mathematical Sciences, Cambridge CB3 0WA, United Kingdom}
\affiliation{TCM Group, Cavendish Laboratory, University of Cambridge, Cambridge CB3 0HE, United Kingdom}

\author{Zhilin Li}
\affiliation{Beijing National Laboratory for Condensed Matter
Physics, Institute of Physics, Chinese Academy of Sciences,
Beijing 100190, China}

\author{Vito Fabio}
\affiliation{Department of Physics, University of Calabria, via
ponte Bucci, 31/C, 87036 Rende (CS), Italy}

\author{Liwei Guo}
\affiliation{Beijing National Laboratory for Condensed Matter
Physics, Institute of Physics, Chinese Academy of Sciences,
Beijing 100190, China}

\author{Xiaolong Chen}
\affiliation{Beijing National Laboratory for Condensed Matter
Physics, Institute of Physics, Chinese Academy of Sciences,
Beijing 100190, China}

\author{Sankar Das Sarma}
\affiliation{Condensed Matter Theory Center and Joint Quantum
Institute, Department of Physics, University of Maryland, College
Park, Maryland 20742-4111 USA}

\author{Antonio Politano}
\email{antonio.politano@iit.it} \affiliation{Istituto Italiano di
Tecnologia, Graphene Labs, Via Morego, 30, 16163 Genova, Italy}
\affiliation{Department of Physical and Chemical Sciences,
University of L'Aquila, Via Vetoio, 67100 L'Aquila, Italy}

\date{\today}

\begin{abstract}
By means of high-resolution electron energy loss spectroscopy, we
investigate the low-energy excitation spectrum of transition-metal
monopnictides hosting Weyl fermions. We observe gapped plasmonic
modes in (001)-oriented surfaces of single crystals of NbAs and
TaAs at 66 and 68 meV, respectively. Our findings are consistent
with theory and we estimate an effective Coulomb
 interaction strength $\alpha_{\rm eff}\approx0.41$ for both samples. We also demonstrate that the
modification of the surface of transition-metal monopnictides by
the adsorption of chemical species (in our case, oxygen and
hydrocarbon fragments) changes the frequency of the plasmonic
excitations, with a subsequent modification of the effective
interaction strength in the 0.30-0.48 range. The remarkable
dependence of plasmonic features on the presence of adsorbates
paves the way for plasmonic sensors based on Weyl semimetals
operating in the mid-infrared.

\end{abstract}

\maketitle

Massless Dirac fermions, previously observed in two dimensions in
graphene~\cite{novoselov05} and topological
insulators~\cite{luo13}, have recently been observed in
three-dimensional topological Dirac semimetals
(TDS)~\cite{borisenko14,liu14,liu14b} and topological Weyl
semimetals (TWS)~\cite{armitage17}. These three-dimensional
semimetals represent a novel state of quantum matter with bulk
Weyl fermions and non-trivial topological surface states forming
Fermi semi-arcs~\cite{belopolski16,moll16}. The topological  bulk
and surface band structure of TWS gives rise to a plethora of
novel phenomena, such as negative magnetoresistance~\cite{son13},
chiral magnetic effects~\cite{song16,cortijo16}, chiral
magnetoresistance~\cite{niemann17,burkov15}, and anisotropic
Adler-Bell-Jackiw anomaly~\cite{lv17}.  The signatures of TWS have
been revealed in NbP~\cite{chang16,liu16},
TaP~\cite{liu16,arnold16,xu15,xu16},
TaAs~\cite{huang15,weng15,lv15b,lv15,yang15,xu15b}, and
NbAs~\cite{xu15c}.

In particular, the use of topological materials has opened new
pathways for plasmonics~\cite{politano17}. As the separation of
Weyl nodes in both momentum and energy acts as an effective
applied magnetic field in momentum space, plasmonic modes in TWS
behave similarly to magnetoplasmons in ordinary
metals~\cite{hofmann16,yao17}. The application of magnetic fields
also activate further intriguing peculiarities, including
coupling-induced transparency and broadband polarization
conversion~\cite{long18}. Therefore, TWS are very interesting
candidates for plasmonic devices.

However, to date the use of TWS in plasmonics has been hindered by
the absence of samples with high crystalline quality.
Specifically, the crystal growth from melt is impeded by the
facile sublimation of As before melting and by the presence of
other phases in the Ta-As~\cite{murray76} and
Nb-As~\cite{warczok12} binary system. Moreover, in standard
conditions, chemical vapor transport using iodine as transport
agent only provides crystals with small size (\numrange{0.5}{1.5}
\SI{}{\milli\metre}). By using foils instead of powder as the
starting material, tilting the ampoule to improve convections, and
adjusting the concentration of agent iodine, we were able to
optimize the growth and obtain single-crystal foils of
transition-metal arsenides in the centimeter scale with
unprecedented crystalline quality~\cite{li16}.

In this Letter, we make use of this experimental advance in
crystal growth to study  the plasmonic spectrum of both NbAs and
TaAs single crystals by means of high-resolution electron energy
loss spectroscopy (HREELS). We find that the plasmon
frequency is similar in both compounds independent of the doping
level, in agreement with predictions from the high-temperature
random phase approximation (RPA). Moreover, we discover that the
plasmon frequency can be tuned when chemical species are adsorbed
at NbAs and TaAs surfaces, highlighting the high potential of TWS
plasmonic modes in the field of sensors.

\begin{figure}[t]
\scalebox{0.65}{\includegraphics{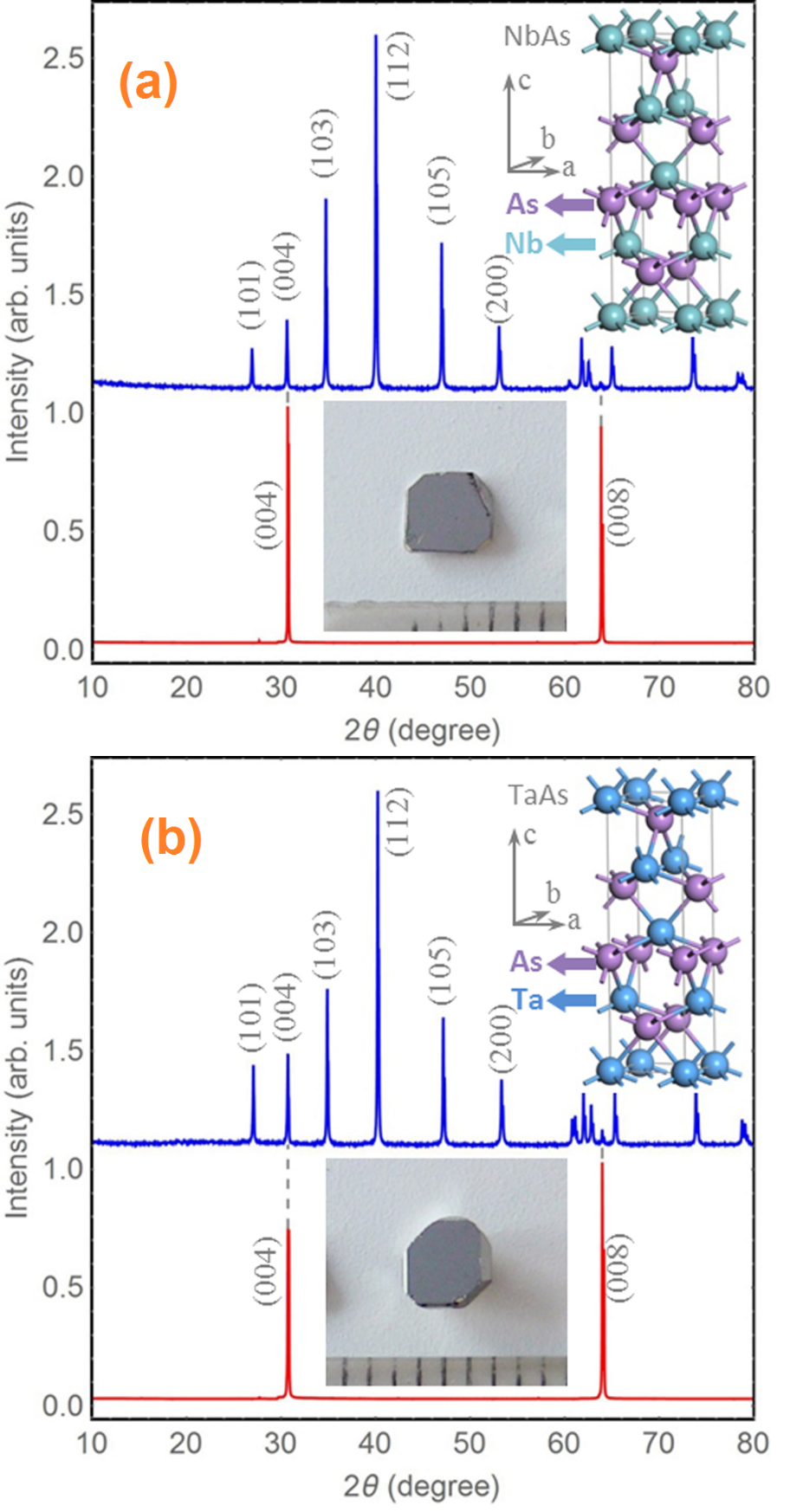}} \caption{(a) XRD
pattern of the as-grown NbAs crystal (CuK$_\alpha$). The top panel
shows the pattern for NbAs powder, while the bottom panel is
related to the (001)-oriented flat surface of the single crystal.
The inset of the top panel shows the crystal structure (purple for
As and blue for Nb). The bottom panel shows the sample compared
with a millimeter ruler. (b) XRD pattern of TaAs (with blue for
Ta). The notation is the same as in (a).} \label{fig:1}
\end{figure}

Our TaAs and NbAs crystals were grown by chemical vapor transport
method~\cite{li16}. In a typical run, tantalum (or niobium),
arsenic, and iodine (transport agent,
\SI{5}{\milli\gram/\milli\litre}) were loaded in a silica ampoule
under argon. The ampoule was then evacuated, sealed, and heated
gradually from room temperature to \SI{1050}{\celsius} over
\SI{72}{} hours. Afterwards, the ampoule was put to a temperature
gradient from \SI{1070}{\celsius} to \SI{1030}{\celsius} for two
weeks and naturally cooled down to room temperature. As shown in
Fig.~\ref{fig:1}, the obtained crystals have regular shapes and
shiny facets. Experiments were performed in an ultra-high vacuum
(UHV) chamber with a base pressure of 10$^{-11}$ mbar. The absence
of contaminants has been ensured by means of Auger electron
spectroscopy (AES) and vibrational spectroscopy, while X-ray
diffraction (XRD) and low-energy electron diffraction (LEED) have
been used to study bulk and surface structural order,
respectively. All experiments have been carried out at room
temperature. Crystals show large-scale terraces (001)-oriented, as
also confirmed by XRD (Fig.~\ref{fig:1}). We chose those surfaces
since the Weyl cones in the bulk and the Fermi-arc surface states
are observed in the (001)-terminated surface on both
NbAs~\cite{xu15c} and TaAs~\cite{huang15}. Samples have been
cleaved in situ in UHV. The experimental crystal structure of NbAs
and TaAs is shown as an inset in Fig.~\ref{fig:1}. In NbAs (TaAs),
Nb (Ta) and As atoms are six-coordinated to each other. They
crystallize in a body-centered tetragonal unit cell with lattice
constants $a = \SI{3.45}{\angstrom}$ (\SI{3.44}{\angstrom}) and $c
= \SI{11.68}{\angstrom}$ (\SI{11.64}{\angstrom}). For both
samples, the space group is I4$_1$md (No. 109) and the structure
lacks inversion symmetry.

\begin{figure}[t]
\scalebox{0.2}{\includegraphics{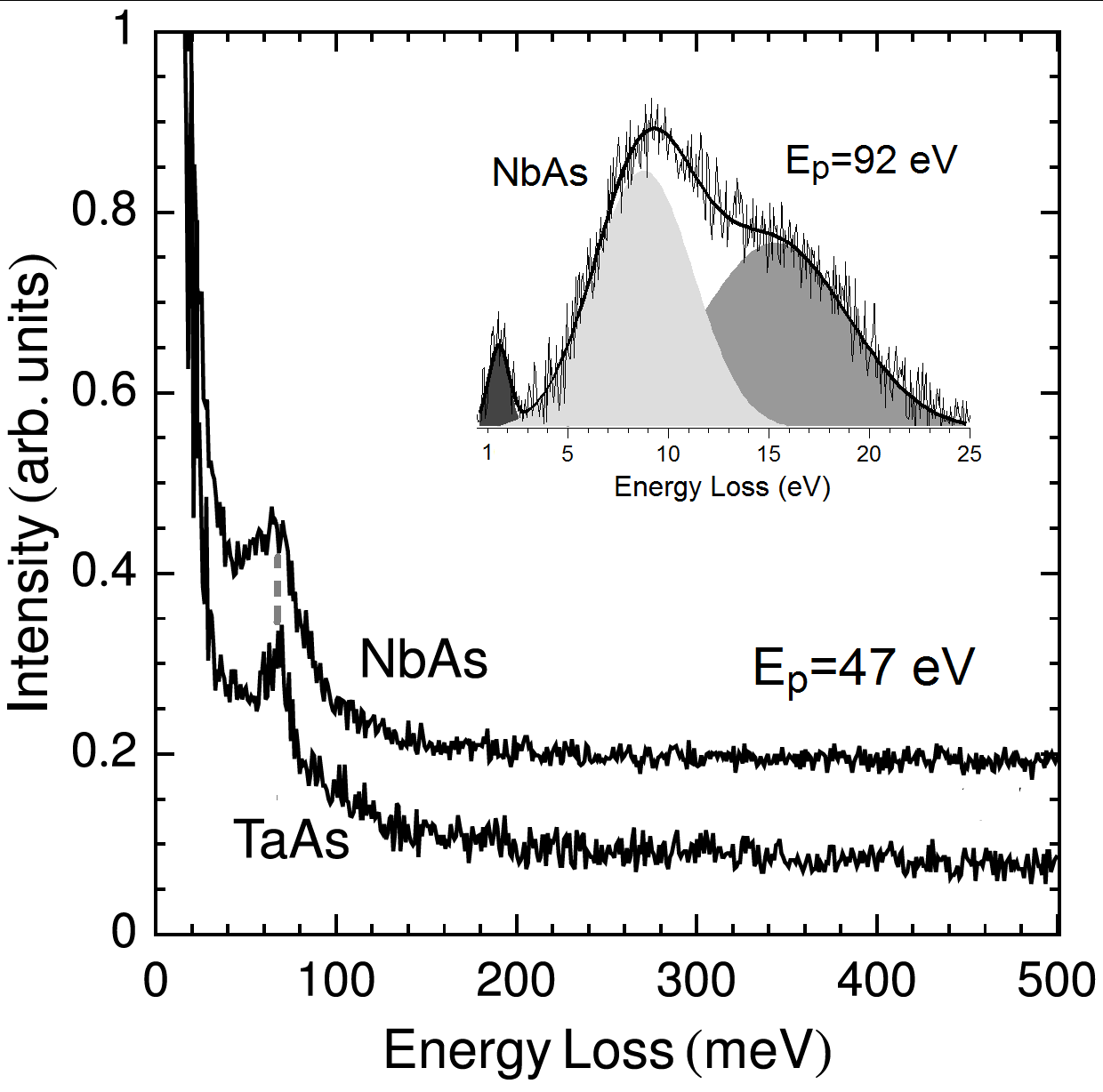}} \caption{HREEL spectra
of (001)-oriented NbAs and TaAs single crystals at room
temperature acquired for an impinging energy of 47 eV in grazing
scattering conditions. The spectra are off-set vertically. The
position of the plasmon resonance is indicated by the vertical
dashed gray line. Note that the excitation spectrum probed by
HREELS in the specular geometry does not exhibit the optical
phonon modes of NbAs and TaAs at
\SI{33}{\milli\eV}~\cite{chang16b} and
\SI{32}{\milli\eV}~\cite{chang16b,liu15}, respectively, since
phonon modes can be excited in HREELS only in the off-specular
(impact scattering) geometry~\cite{dejuan15} The inset shows the
excitation spectrum in an extended energy range probed for
NbAs(001) for an impinging energy of 92 eV.} \label{fig:2}
\end{figure}

The HREELS excitation spectrum of (001)-oriented pristine NbAs and
TaAs single crystals at room temperature in shown in
Fig.~\ref{fig:2}. The experiments were performed with a Delta 0.5
spectrometer by Specs GmbH, Germany, in a specular geometry, i.e.,
with equal incident and scattered electron beam angles
$\alpha_i=\alpha_s=72^{\circ}$ with respect to the surface normal,
with an incident energy $E_i=\SI{47}{\eV}$ in order to minimize
the inelastic mean free path for electrons~\cite{da14} and to increase the surface sensitivity by
enhancing the spectral weight of surface-related excitations. The
loss spectra of both samples exhibit a single peak at a similar
energy at \SI{66}{\milli\eV} and \SI{68}{\milli\eV}, respectively.
The angle-integrated scattering cross section for
HREELS~\cite{mills75,ibach82}
\begin{align}
\frac{d\sigma}{dE} &\sim \frac{1+n(E)}{E \cos \alpha_i} \, {\rm Im} \bigl[1+\varepsilon(E, q_\parallel)\bigr]^{-1} , \label{eq:crosssection}
\end{align}
has a resonance for surface plasmonic modes $1+\varepsilon(E,
q_\parallel) = 0$~\cite{mills75,ibach82}. Here, $E$ is the energy
loss transferred to the surface, $n(E)$ the Fermi-Dirac
distribution, and $q_\parallel$ is the parallel momentum
transfer~\cite{rocca95}:
\begin{align}
q_\parallel &= \frac{\sqrt{2mE_i}}{\hbar} \bigl(\sin \alpha_i -
\sqrt{1-\frac{E}{E_i}} \sin \alpha_s \bigr).
\end{align}
Hence, we identify the peaks at \SI{66}{\milli\eV} and
\SI{68}{\milli\eV} with the surface plasmon modes of the TWS. The
plasmon frequency is quite similar in both materials, indicating
that the Fermi velocity and the interaction strength are of
comparable magnitude, consistent with band-structure
calculations~\cite{lee15}. Recent optical conductivity measurement
on the possible TWS Eu$_2$Ir$_2$O$_7$~\cite{sushkov15} reported
plasmon energies in the $\numrange{50}{90}\, \SI{}{\milli\eV}$
range, quite consistent with our findings for NbAs and TaAs. It is
worthwhile pointing out that the observed excitations cannot be
associated to phonons or polaritons. As a matter of fact,
composite modes arising from the hybridization of phonon modes
with plasmons should have a gap of the order of the highest branch
of optical phonons in NbAs (33 meV~\cite{chang16b}) and TaAs (32
meV~\cite{chang16b,liu15}). In the excitation spectrum acquired
with a primary electron beam energy of 92 eV (inset of Fig. 2, for
the case of NbAs), we observe also other resonances at ~\SI{1.5},
~\SI{8.8} and ~\SI{15.3} {\eV} (inset of Fig.~\ref{fig:2}). While
the first two features are originated by interband transitions,
the latter excitation can be attributed to the plasmonic response
of large hole pockets~\cite{arnold16b}, consistent with
calculations of the plasma edge~\cite{grassano18}.

The similar values for the plasmon frequency in both samples and
the smallness of typical doping levels (${\sim\SI{10}{\milli\eV}
\approx \SI{100}{\kelvin}}$, close to the charge-neutral
point~\cite{arnold16b}) indicate that we probe the universal
high-temperature regime, where the plasmon is formed of thermally
excited interband carriers and which is independent of the doping
level~\cite{hofmann15}.  Within the RPA, the dielectric function
at finite temperature and doping was computed in
Ref.~\cite{hofmann15}. At high temperature, an analytical
expression is given by~\cite{hofmann15}
\begin{align}
E_{\rm plas} = \sqrt{\frac{2\pi g \alpha}{\kappa(T)}} \frac{T}{3} ,
\label{eq:plasmonrpa}
\end{align}
where $T$ is the temperature (room temperature in our case), $v_F$
is the Fermi velocity, $g$ is the number of Weyl nodes ($g = 24$
here) and $\alpha = e^2/\hbar v_F \varepsilon_\infty$ is the
dimensionless Coulomb interaction strength (the WSM ``fine
structure constant'', in analogy with quantum
electrodynamics~\cite{throckmorton15}), usually ranging between
0.2 and 2. In addition, $\kappa(T)$ accounts for the effective
temperature-dependent electronic renormalization of the dielectric
constant, commonly parametrized in terms of a ``Landau pole''
scale $\Lambda_L$ as $\kappa(T) = \frac{g\alpha}{3\pi} \ln
\frac{\hbar v_F \Lambda_L}{\pi T e^{-\gamma_E}}$, where $\gamma_E$
is Euler's constant~\cite{hofmann15}. The fact that the
long-wavelength plasma frequency~\eqref{eq:plasmonrpa} of WSM
depends on the interaction coupling constant and the Landau scale
is a direct manifestation of the non-Fermi-liquid nature of
intrinsic WSM by virtue of the Weyl point being a quantum critical
point~\cite{throckmorton18}.

The appearance of the Landau pole is a manifestation of electron
interactions in the WSM: while the classical theory of a WSM is
scale-invariant (i.e., the single-particle Dirac Hamiltonian
scales in the same way as the Coulomb interaction under a change
of length scales), this classical symmetry is broken once quantum
effects through the Coulomb interaction are taken into account.
The dimensionless coupling $\alpha$ of the classical system now
depends on the temperature or doping scale at which it is
measured. RPA predicts this ``running'' of the coupling from weak
coupling at small temperatures $T\ll \Lambda_L$ to strong coupling
as $T$ approaches the Landau scale $\Lambda_L$, where the RPA
breaks down~\cite{throckmorton15}. The dimensionless coupling
constant $\alpha$ is replaced by the dimensional quantity
$\Lambda_T$, which is known as ``dimensional transmutation''. For
the plasmon~\eqref{eq:plasmonrpa}, dimensional transmutation
implies that the naive (by dimensional analysis)
linear-in-temperature dependence of the frequency is replaced by a
super-linear temperature-scaling that depends on the Landau pole.
A fit of the Landau pole (assuming the RPA result) has predictive
power for other temperature and density scales, whereas the  fine
structure constant $\alpha(T)$ is always measured at a fixed
temperature.

Both TaAs and NbAs have a total of 24 Weyl nodes, which are
divided in two classes W1 and W2 with different band parameters.
We fit the peaks in Fig.~\ref{fig:2} using the high-temperature
expression with $T = 300 K$. The fit value for the Landau pole is
$\Lambda_L = 126\pm35 \SI{}{\milli\eV}$ and $\Lambda_L = 118\pm18
\SI{}{\milli\eV}$ for NbAs and TaAs, respectively, corresponding
to an effective interaction strength of $\alpha_{\rm eff} =
\frac{\alpha}{\kappa(T)} = 0.41\pm0.07$. This large value of the
interaction strength (at room temperature) indicates that the WSMs
TaAs and NbAs are not weakly coupled and are no longer described
by leading-order perturbation theory in $\alpha$
~\cite{throckmorton15}.

Most practical application of TWS should work in ambient
conditions. Thus, it is crucial to evaluate the effect of surface
modifications on the plasmon frequency. In addition, tunability of
surface-plasmon frequency is highly desirable for the prospect of
plasmonic sensors~\cite{sturaro16,nasir14}.

\begin{figure}[t]
\scalebox{0.85}{\includegraphics{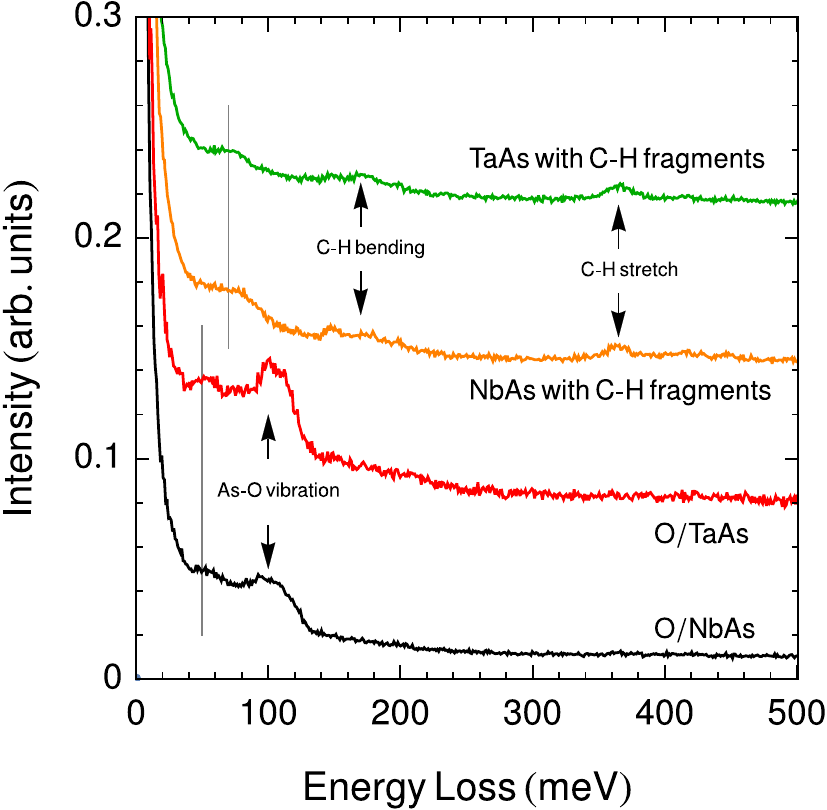}} \caption{HREEL
spectrum of TaAs and NbAs with C-H fragments (top two solid green
and orange lines) and O-modified surfaces (bottom two solid red
and black lines). The spectra are off-set vertically. The vertical
solid gray lines denote the position of the plasmon resonance,
which is red-shifted in O-modified samples compared to the
pristine surface, and blue-shifted for samples modified by the
adsorption of hydrocarbon fragments. Loss spectra show additional
features associated with C-H bending and stretching vibrations and
As-O vibrational modes, respectively, which are indicated by the
arrows.} \label{fig:3}
\end{figure}

Here, we consider two case-study examples of chemisorbed species
providing opposite sign in the shift in the surface-plasmon
frequency. Figure~\ref{fig:3} shows the behavior of plasmonic
modes in TWS samples saturated with oxygen at room temperature
(bottom two black and red lines). Being As-terminated, TWS samples
are not inert and interact strongly with oxygen~\cite{politano18}. The resulting loss spectrum is dominated by a
band centered at \SI{100}{\milli\eV}~\cite{silva13}, with the main
contribution from the As-O stretching mode and minor spectral
contribution from Nb(Ta)-O stretching~\cite{politano18}. The plasmon frequency is red-shifted compared to
the pristine crystal to \SI{58}{\milli\eV} in both NbAs and TaAs
(solid gray line). This red-shift can be accounted for by a change
in the surface dielectric function. The plasmon condition changes
to $\varepsilon_s(E) + \varepsilon(E) = 0$, where the dielectric
function $\varepsilon_s(E)$ due to adsorbed particles
becomes~\cite{ibach82}
\begin{align}
\varepsilon_s(E) = 1 + \frac{4\pi {e^*}^2 n}{M} \frac{1}{E_T^2 - E^2 - i E \Gamma} ,
\end{align}
where $E_T$ is the bare oscillation frequency of the As-O bond.
The adsorbate contributes to the dielectric function with a
harmonic oscillator term, with a small (though existing)
hybridization between surface collective electronic excitations
and the adsorbate mode. Thus, the red-shift can be accounted for
by a coupling of the adsorbate layer and the plasmonic mode as:
\begin{align}
\varepsilon(E) = \varepsilon_\infty \biggl(1- \frac{E_p^2}{E (E + i\Gamma)}\biggr) + \Delta \varepsilon \frac{E_T^2}{E_T^2 - E^2 - i E \Gamma'} .
\end{align}
A fit to the data gives $\Delta\varepsilon = 8$.
$\varepsilon_s(E)$ is positive for $E < E_T$ and thus decreases
the plasmon  frequency. The effective interaction strength is
$\alpha = 0.30\pm0.15$ and $\alpha = 0.31\pm0.16$ for O-modified
NbAs and TaAs, respectively. The Landau pole $\Lambda_L$ in the
two cases is $165\pm108$ and $165\pm112$ meV.

It is also worth mentioning that the shift of the plasmon
frequency due to changes in carrier density arising from
adsorbate-induced surface doping further supports the
interpretation of the modes in pristine NbAs and TaAs
(Fig.~\ref{fig:1}) as surface rather than bulk excitations.
Especially, interaction with oxygen has been recently demonstrated
only to affect Fermi-arc states~\cite{politano18}, while the O-induced modification of the other electronic
bands is insignificant.

An opposite change in the plasmon frequency (different from the
plasmon-adsorbate hybridization) can arise from an effective
charge transfer from adsorbates to the system or vice versa. Such
a doping effect plays out when hydrocarbon fragments are present
over the surface of the TWS. Their presence is evident in HREELS
spectra in Fig.~\ref{fig:3} (top two orange and green lines),
which display C-H bending and stretching modes
at~\SI{170}{\milli\eV} and \SI{360}{\milli\eV},
respectively~\cite{sakong10,politano13}, arising from the
decomposition of hydrocarbons over the surfaces of NbAs and TaAs.
The plasmon frequency is blue-shifted to \SI{72}{\milli\eV} and
\SI{73}{\milli\eV} in NbAs and TaAs, respectively. We ascribe this
shift to the electron transfer (n-doping) from C-H groups to NbAs
and TaAs. The fit gives an effective interaction strength $\alpha
= 0.48\pm0.08$ and $\alpha = 0.47\pm0.21$ for NbAs and TaAs,
respectively. The Landau pole $\Lambda_L$ in the two cases is
$103\pm14$ and $106\pm40$ meV.

Finally, in order to evaluate the feasibility of TWS-based
plasmonics it is important to assess the lifetime of the plasmonic
modes through the analysis of the line-width of the HREEL peak. The lifetime of surface collective electronic excitations
notably influences plasmon-mediated dynamic processes, the field
enhancement of surface enhanced Raman scattering
(SERS)~\cite{kneipp97}, and surface-enhanced
fluorescence~\cite{lakowicz04}. In this case, we only note a
decrease of the lifetime of the plasmonic excitation by less than
one half, as chemisorbed species favor phase-breaking scattering
events through collisions of electrons with subsequent enhancement
of the Landau damping and, consequently, a broadening of the
plasmonic peak. However, this increase of the lifetime is
relatively low compared to the total quenching induced by
chemisorbed species experienced for the case of graphene plasmons~\cite{politano13b}, in spite of the low chemical reactivity of
graphene~\cite{qi11}.

In conclusion, we have studied the low-energy excitation spectrum
in (001)-oriented NbAs and TaAs surfaces. The loss spectrum
measured by HREELS, dominated by surface plasmons in the
mid-infrared, has been fitted by RPA model in order to extract
both the effective Coulomb interaction strength and the Landau
pole. The pristine NbAs and TaAs single crystals are characterized
by a single plasmonic resonance at \SI{66}{\milli\eV} and
\SI{68}{\milli\eV}, respectively. The strong reactivity toward
oxygen has noticeable implications on the plasmonic spectrum of
TWS, due to the change of the dielectric function induced by
oxygen adatoms stably adsorbed at room temperature. The plasmon
energy is shifted down to \SI{58}{\milli\eV} for both NbAs and
TaAs, with a subsequent reduction of the effective strength. In
the presence of fragments of hydrocarbons, a charge transfer from
adsorbates to TWS occurs with a blue-shift of the plasmonic
frequency up to 72 and \SI{73}{\milli\eV} in NbAs and TaAs,
respectively. Our results evidence that the plasmonic features of
the pristine NbAs and TaAs surfaces can be preserved only by
protecting the active channel of the TWS-based plasmonic device
with a capping layer in order to reduce interaction with
environmental species. On the other hand, the striking dependence
of plasmonic features of TWS on the presence of adsorbates can be
used to develop a new generation of plasmonic chemical sensors.

\begin{acknowledgments}

JH acknowledges support by Gonville and Caius College and Peterhouse, Cambridge.
\end{acknowledgments}

\bibliography{bib}

\end{document}